# Rotating surface soliton complexes in annular waveguides

Zhiyong Xu

*Nonlinear Physics Centre and CUDOS, Research School of Physical Sciences and Engineering, the Australian National University, Canberra ACT, 0200, Australia &*
*ICFO-Institut de Ciencies Fotoniques, and Universitat Politecnica de Catalunya, Mediterranean Technology Park, 08860 Castelldefels (Barcelona), Spain*

Email: *xzy124@rsphysse.anu.edu.au*

Abstract: We address surface soliton complexes formed at the edge of annular guiding structures containing several concentric rings. Such soliton complexes feature a $\pi$-phase difference between neighboring spots. It is shown that the multipole-mode solitons can rotate steadily upon propagation, and the existence domain is strongly affected by the rotation frequency. The rotation may enhance the stabilization of surface multipole-mode solitons.

PACS: 42.65.Tg, 42.65.Jx, 42.65.Wi

Keywords: Surface waves, Nonlinear waveguide, Soliton

The study of light propagating at the interface between different materials is a fundamental problem of scientific and practical importance [1-4]. Such interfaces can support localized waves, i.e surface states, which were encountered in various areas of physics. In optics, nonlinear surface waves have been investigated extensively during the last few years [5-8]. However, most of these works are focused on one-dimensional (1D) geometry. One natural question is whether the concept of 1D surface solitons can be extended to two-dimensional (2D) settings. In 2006, it was suggested that the edges and corners of finite two-dimensional lattices could support surface solitons [9]. Very recently, this idea was verified by two remarkable experiments, which demonstrated 2D surface solitons in optically induced lattices in photorefractive crystal [10] and fs-laser-written arrays in bulk fused silica [11], respectively.

Soliton dynamics depends crucially on the overall geometry of the lattice. In this respect the lattices with radial symmetry afford the wealth of possibilities for soliton control [12, 13], including the possibility to study mobility of solitons in experiment. In the circular geometry, in particular, the structures created by Bessel beams support rotary solitons, stable vortices in self-defocusing media, and multipole-mode solitons [14-17]. Note that solitons in the ring structure have been demonstrated in [18]. Very recently, surface solitons and switching with vortex beams have been discussed in circular guiding structures with limited number of rings [19, 20].

The aim of this paper is to investigate the properties and stability of multipole-mode solitons residing in the outer ring of guiding structure consisting of several concentric rings. In particular, it will be shown that such surface multipole modes can rotate upon propagation. It is important to note that rotation may help to stabilize multipole-mode surface solitons.

We consider the propagation of a light beam along the $\xi$ axis in a Kerr-type medium with a transverse modulation of refractive index, described by the following equation for the complex field amplitude $q$:

$$i\frac{\partial q}{\partial \xi} = -\frac{1}{2}(\frac{\partial^2 q}{\partial \eta^2} + \frac{\partial^2 q}{\partial \varsigma^2}) - |q|^2 q - pR(\eta,\varsigma)q, \qquad (1)$$

where the transverse $\eta,\varsigma$ and longitudinal $\xi$ coordinates are scaled to the beam width and the diffraction length, respectively; $p$ stands for the dimensionless lattice depth and the function $R(\eta,\varsigma)$ describes the transverse refractive index modulation in the guiding structure with several concentric rings. Thus, we set $R = \cos^2(\Omega r)$ for $r \leq r_{\text{out}}$ and $R = 0$ for $r > r_{\text{out}}$, where $r^2 = \eta^2 + \varsigma^2$ is the radius, and $r_{\text{out}} = (2n-1)\pi/2\Omega$ being the outer radius of the annular waveguide, $\Omega$ being the radial frequency, and $n$ sets the number of rings in the structure. Figure 1(a) shows an example of such structure for $n = 4$. Here we set $\Omega = 2$. It should be pointed out that the truncated annular waveguides considered here are substantially different from radially symmetric Bessel lattices, where the amplitudes and widths of rings constituting the lattice decrease with an increase of radial variable [14-16]. It is also interesting to compare our results with those obtained for guiding structures with infinite number of rings that were investigated in [12, 13]. The difference between properties of multipole-mode solitons in our settings and solutions obtained in the above mentioned Bessel and infinite ring lattices will be discussed below in details. We stress that the guiding structure considered in this paper is *truncated*. This introduces specific features into properties of solitons considered here and is allowed to discuss the concept of *surface* states. The structure that we address here is reminiscent to Bragg and Omniguide fibers [21], which have been fabricated technologically, and are in commercial use nowadays. Another possible way to generate our setting is to use potential optical induction, for instant, a set of concentric rings in photorefractive nonlinear crystal can be induced by using nondiffracting linear optical beams with cylindrical symmetry [22].

We are interested in surface solitons that may rotate, thus the stationary solutions of Eq. (1) were searched in the form $q(\eta,\varsigma,\xi) = [u(\eta',\varsigma') + iv(\eta',\varsigma')]\exp(ib\xi)$ ($u,v$ are the real and imaginary parts of the field amplitude, and $b$ is the propagation constant) in the rotating coordinate frame defined as $\eta' = \eta\cos(\alpha\xi) + \varsigma\sin(\alpha\xi)$ and $\varsigma' = \varsigma\cos(\alpha\xi) - \eta\sin(\alpha\xi)$, where $\alpha$ is the rotation frequency. In these coordinate, the equations for the filed components take the following form (for simplicity, the prime coordinates are omitted further):

$$\alpha\eta\frac{\partial v}{\partial \varsigma} - \alpha\varsigma\frac{\partial v}{\partial \eta} + \frac{1}{2}\left(\frac{\partial^2 u}{\partial \eta^2} + \frac{\partial^2 u}{\partial \varsigma^2}\right) + u(u^2+v^2) + pRu = bu,$$
$$\alpha\varsigma\frac{\partial u}{\partial \eta} - \alpha\eta\frac{\partial u}{\partial \varsigma} + \frac{1}{2}\left(\frac{\partial^2 v}{\partial \eta^2} + \frac{\partial^2 v}{\partial \varsigma^2}\right) + v(u^2+v^2) + pRv = bv, \qquad (2)$$

Equation (1) conserves the power, Hamiltonian and angular momentum, among which the power is $U = \iint_{\infty} |q|^2 \mathrm{d}\eta \mathrm{d}\varsigma$. The system (2) is solved numerically by Newton relaxation method.

The system (1) supports fundamental solitons (monopoles) residing in the outer ring of the truncated annular waveguide. We found that the power of such solitons is a nonmonotonic function of the propagation constant $b$ at $\alpha = 0$ and there exists a lower cutoff of $b$ (just as in the case of surface solitons at the edge of periodic lattices), while above a certain rotating frequency the power of solitons becomes a monotonic function of $b$ and it vanishes at the cutoff of the propagation constant. Thus the salient point is that rotating monopoles do not feature power threshold for their formation if the rotation frequency is high enough [19].

The main objective of this work is to consider properties of multipole-mode surface waves, which feature several amplitude modulations and a $\pi-$ phase difference between neighboring bright spots forming such waves. The typical profile of simplest surface wave of this type, i.e. surface dipole-mode soliton is shown in Fig. 1(b) for the case of zero rotation frequency. The power of such surface mode is a nonmonotonic function of the propagation constant [Fig. 2(a)]. At high powers, the dipole-mode soliton resembles a pair of narrow almost noninteracting bright surface solitons, however, at lower powers dipole-mode solitons tend to extend deep into the ring guiding structure. There exists a lower cutoff on the propagation constant for such solitons.

Once soliton rotation takes place, the properties of surface multipole-mode solitons change dramatically. As shown in Fig. 2(a), the power of rotating dipole-mode solitons becomes monotonic function of the propagation constant when rotation frequency exceeds a certain minimal value [Fig. 2(a)]. At low powers (i.e. near the cutoff on propagation constant), rotating dipole-mode solitons show a clear tendency to expand not only on the radial direction but also in the azimuthal direction. Far from the cutoff such solitons are well localized in both radial and azimuthal directions [Fig. 1(d)]. We found that the cutoff on propagation constant $b_{\mathrm{co}}$ increases monotonically with an increase of the rotation frequency [Fig. 3(a)]. The larger the difference $b_{\mathrm{co}}(\alpha) - b_{\mathrm{co}}(0)$, the stronger the soliton is localized along the radial coordinate, even when the propagation constant approaches the cutoff. It should be stressed that the power of dipole-mode solitons does not vanish completely in the cutoff on the propagation constant, while instead in this point multiple-mode solitons transform into ring-shaped solutions, in a sharp contrast with the fundamental solitons where the power vanishes in the cutoff. It should be pointed out that ring-shaped solitons are unstable in this setting, and will break into several fundamental solitons (the details on switching of vortex-type solitons in such an angular waveguide can be found in ref. [20]).

An important finding is that rotating multipole-mode solitons with a fixed propagation constant $b$ cease to exist when the rotation frequency reaches a certain maximal possible value. Thus, for sufficiently large propagation constant ($b > b_{\mathrm{co}}(0)$), the total power of solitons monotonically decreases with an increase of the rotation frequency $\alpha$ and reaches the minimal value when $\alpha \to \alpha_m$ [Fig. 2(b)]. At $\alpha \to \alpha_m$, rotating surface dipole-mode solitons strongly expand along the azimuthal direction and deform into ring-shaped beams. We found that the maximal rotation frequency increases monotonically with $b$ and that it vanishes when $b$ approaches the cutoff on propagation constant for nonrotating dipole soliton [Fig. 3(b)]. Note that changing lattice depth $p$ does not result in qualitative modification on properties of solitons, as illustrated in Figs. 3(a,b). One can see that the dependencies $b_{co}(\alpha)$ and $\alpha_m(b)$ are similar for different values of lattice depth $p$.

We also found more complex multipole-mode solitons residing at the edge of truncated annular guiding structure. Figures 1(e,f) show an example of surface quadrupole-mode solitons, whose salient properties are similar to those of surface dipole-mode solitons.

To elucidate the stability of the whole family of multipole-mode surface solitons, we searched for the perturbed solutions of Eq. (1) in the form $q = [u + iv + w_r + iw_i]\exp(ib\xi)$, where $w_r(\eta,\varsigma,\xi)$ and $w_i(\eta,\varsigma,\xi)$ are the real and imaginary parts of a small perturbation that can grow with a complex rate upon propagation. Linearization of Eq. (1) around stationary solutions obtained from Eqs. (2), at the first order of perturbation theory, results in the following eigenvalue problem

$$\begin{aligned}\frac{\partial w_i}{\partial \xi} &= (3u^2 + v^2)w_r + 2uvw_i + \frac{1}{2}\left(\frac{\partial^2}{\partial \eta^2} + \frac{\partial^2}{\partial \varsigma^2}\right)w_r + pRw_r - bw_r,\\ \frac{\partial w_r}{\partial \xi} &= -(3v^2 + u^2)w_i - 2uvw_r - \frac{1}{2}\left(\frac{\partial^2}{\partial \eta^2} + \frac{\partial^2}{\partial \varsigma^2}\right)w_i - pRw_i + bw_i,\end{aligned} \quad (3)$$

where we assumed that $w_{r,i} \sim \exp(\delta\xi)$. We solved the system (3) numerically in order to find the growth rates $\delta$. To confirm the results of linear stability analysis, we performed the extensive numerical simulations of Eq. (1) with the input condition $q|_{\xi=0} = (u + iv)(1 + \rho)$, where $\rho(\eta,\varsigma)$ are random functions with variance $\sigma_{noise}^2 = 0.01$. The Eq. (1) was solved by using the split-step Fourier method. We typically used transverse meshes with $256 \times 256$ points, transverse steps of the order of $0.05$ and longitudinal steps of the order of $10^{-3}$. The accuracy of the results was checked by doubling the number of points in transverse direction, taking larger integration windows, and by reducing longitudinal steps.

The linear stability analysis shows that for nonrotating surface dipole-mode solitons there exists a narrow instability domain near the cutoff on propagation constant, but above a certain value of propagation constant such solitons become completely stable [Fig. 3(c)]. It should be stressed that near the cutoff on propagation constant the instability develops exponentially, but there also exist an interval of propagation constants where dipoles suffer from oscillatory instability. Direct simulations of Eq. (1) confirmed the results of linear stability analysis.

One of the most important results of this work is that the rotation enhances the stabilization of surface multipole-mode solitons. At small rotation frequencies there still exists a very narrow instability band for surface dipole-mode solitons located close to the cutoff on propagation constant. The perturbed dipole-mode solitons from this band usually evolve into fundamental solitons that rotate along the outer ring of the annular waveguide [Fig. 4]. However, for rotation frequencies exceeding the value $\alpha \approx 0.15$ dipole-mode solitons become completely stable in the whole domain of their existence. A stable rotating dynamics of surface dipole-mode soliton is shown in Fig. 5. Rotation-induced stabilization is even more pronounced for quadrupole-mode solitons. Thus nonrotating quadrupole-mode solitons appear to be unstable (at least for the same parameter values as we took for dipole-mode solitons), while rotating quadrupole-mode solitons can be stable. Fig. 6 shows an illustrative example of stable propagation of rotating quadrupole-mode solitons in the presence of the broadband noise. One can clearly see that such rotating surface multipole-mode solitons can survive over the huge distances. Note that quadrupole-mode solitons are less stable than dipole ones, and higher-order solitons could be stabilized too, thus in general, the more humps the solitons have, the less stable.

We also compared our results with those in infinite radially periodic guiding structures, which are studied in particular in ref. [13]. Some results are presented with dashed curves in Fig. 2. One immediately finds that the threshold power for the formation of multipole-mode solitons in infinite guiding structures is substantially higher than the threshold power in truncated guiding structure [dashed curves in Fig. 2(c)]. The instability domain for nonrotating solitons is much broader in infinite structure [dotted curve in Fig. 2(c) for $\alpha = 0$ ]. At high powers the $U(b)$ dependencies are almost similar for infinite and finite structures, which is natural since well-localized solitons do not feel refractive index modulation in adjacent waveguide channels. In infinite guiding structures the rotation also plays an important role for stabilization of multipole-mode solitons. For example, in such structures dipole-mode solitons become completely stable in the whole domain of their existence at $\alpha = 0.2$ [Fig. 2(c)]. However, in the presence of surface, multipole-mode solitons survive at substantially higher rotation frequency [dashed curve in Fig. 2(b)]. This is because in the case of truncated waveguide the surface helps to confine radiation in radial direction and counteracts the centrifugal force. However, in the infinite annular waveguide if the rotation is too fast (which corresponds to the higher rotation frequency), the centrifugal force acting on the soliton gives rise to the power leaking from the circulating soliton into the adjacent radial rings, which leads to disintegration of the soliton [13].

To summarize, we have found that the edge of the annular guiding structure consisting of several concentric rings support multipole-mode solitons. We showed that the existence domain and properties of such multipole-mode solitons are strongly affected by their rotation frequency. We found that the rotation enhances the stabilization of surface multipole-mode solitons. We also discussed the difference between solitons in our system and in guiding structures with infinite number of rings. The results suggest new ways to manipulate light beams.

**Acknowledgements**

The author acknowledges fruitful discussions with Prof. Lluis Torner, Dr. Yaroslav. Kartashov, and Prof. Yuri Kivshar. This work has been supported in part by the Australian Research Council through the Federation Fellowship research projects, and the Generalitat de Catalunya, Spain.

**Figure captions**

**Figure 1**. (a) Profile of ring-like guiding structure. Field modulus distributions for dipole-mode solitons at (b) $b=6.42$, $\alpha=0$, (c) $b=6.6$, $\alpha=0.2$, (d) $b=7.0$, $\alpha=0.2$, and for quadrupole-mode solitons at (e) $b=6.6$, $\alpha=0.2$, and (f) $b=7.3$, $\alpha=0.2$. White circles indicate the center of the outer ring of annular waveguide. In all cases $p=10$, $\Omega=2$, and $n=4$.

**Figure 2**. Power of dipole-mode soliton versus (a) propagation constant and (b) rotation frequency. Points marked by circles in (a) correspond to solitons shown in Fig. 1. (c) Power versus propagation constant for dipole-mode solitons in finite (solid curve) and infinite annular (dashed curve) waveguides, where unstable region is marked by dotted curves. In all cases $p=10$, $\Omega=2$, and for finite waveguide with $n=4$.

**Figure 3**. (a) Cutoff on propagation constant versus rotation frequency, and (b) maximal rotation frequency versus propagation constant of dipole-mode solitons for different $p$. (c) Real part of perturbation growth rate versus propagation constant at $\alpha=0$ and $p=10$. In all cases, $\Omega=2$, and $n=4$.

**Figure 4**. Evolution of an unstable dipole-mode soliton rotating with frequency $\alpha=0.1$ at $p=10$, $\Omega=2$, and $n=4$.

**Figure 5**. Stable propagation of a perturbed dipole-mode soliton rotating with frequency $\alpha=0.15$ at $p=10$, $\Omega=2$, and $n=4$.

**Figure 6**. Stable propagation of a perturbed quadrupole-mode soliton with frequency $\alpha=0.1$ at $p=10$, $\Omega=2$, and $n=4$.

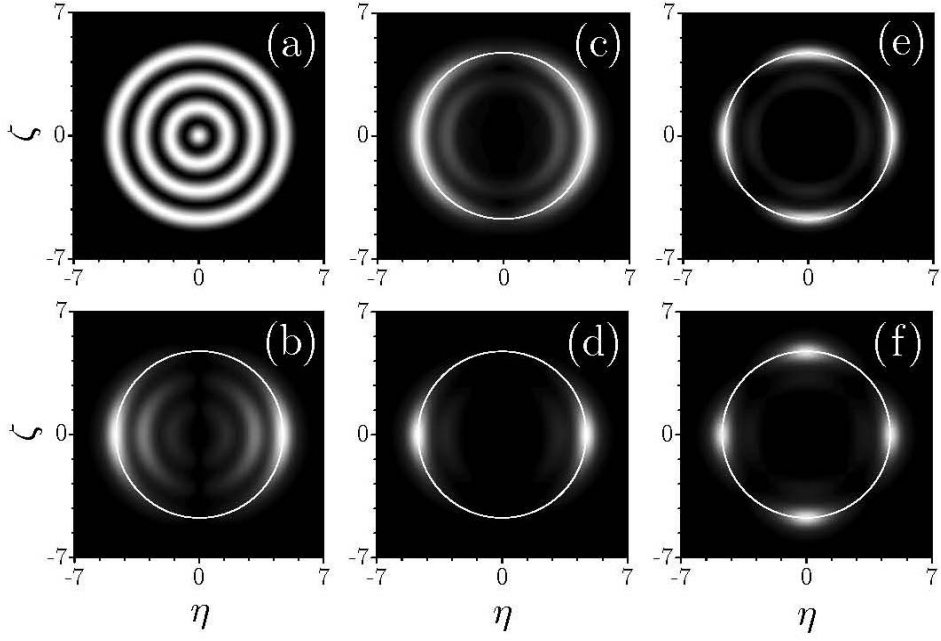

**Figure 1**. (a) Profile of ring-like guiding structure. Field modulus distributions for dipole-mode solitons at (b) $b = 6.42$, $\alpha = 0$, (c) $b = 6.6$, $\alpha = 0.2$, (d) $b = 7.0$, $\alpha = 0.2$, and for quadrupole-mode solitons at (e) $b = 6.6$, $\alpha = 0.2$, and (f) $b = 7.3$, $\alpha = 0.2$. White circles indicate the center of the outer ring of annular waveguide. In all cases $p = 10$, $\Omega = 2$, and $n = 4$.

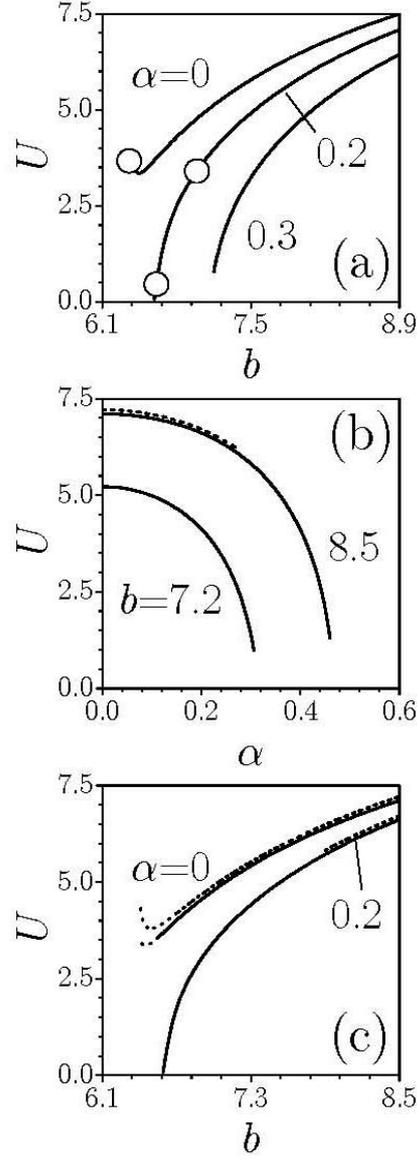

**Figure 2**. Power of dipole-mode soliton versus (a) propagation constant and (b) rotation frequency. Points marked by circles in (a) correspond to solitons shown in Fig. 1. (c) Power versus propagation constant for dipole-mode solitons in finite (solid curve) and infinite annular (dashed curve) waveguides, where unstable region is marked by dotted curves. In all cases $p=10$, $\Omega=2$, and for finite waveguide with $n=4$.

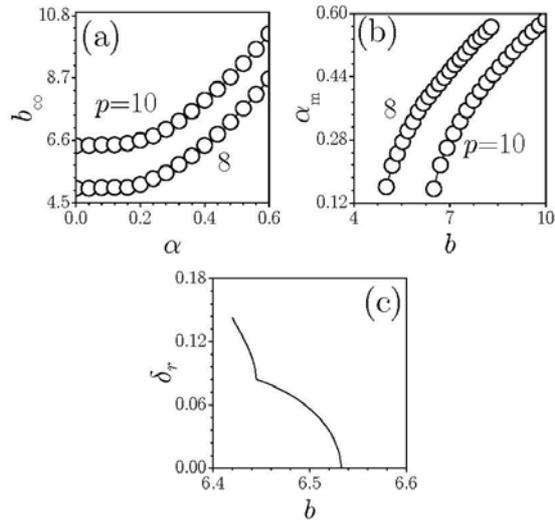

**Figure 3**. (a) Cutoff of propagation constant versus rotation frequency, and (b) maximal rotation frequency versus propagation constant of dipole-mode solitons for different $p$. (c) Real part of perturbation growth rate versus propagation constant at $\alpha = 0$ and $p = 10$. In all cases, $\Omega = 2$, and $n = 4$.

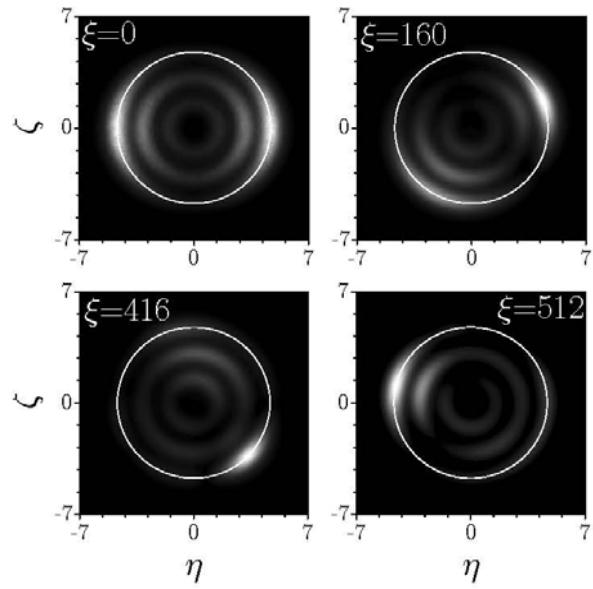

**Figure 4**. Evolution of an unstable dipole-mode soliton rotating with frequency $\alpha = 0.1$ at $p = 10$, $\Omega = 2$, and $n = 4$.

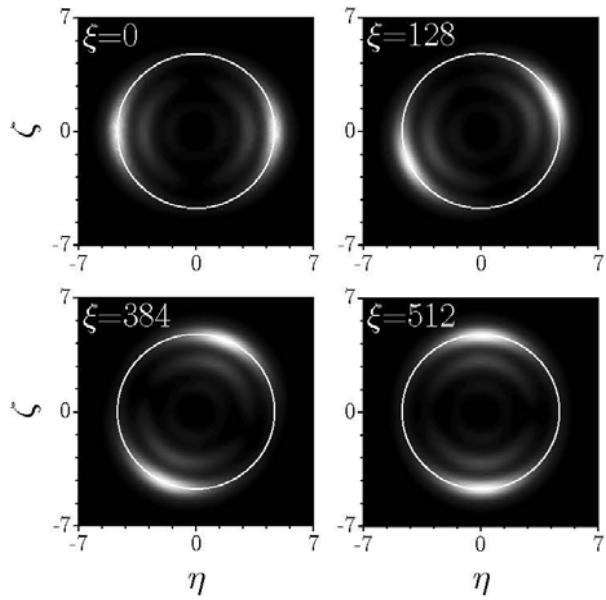

**Figure 5**. Stable propagation of a perturbed dipole-mode soliton rotating with frequency $\alpha = 0.15$ at $p = 10$, $\Omega = 2$, and $n = 4$.

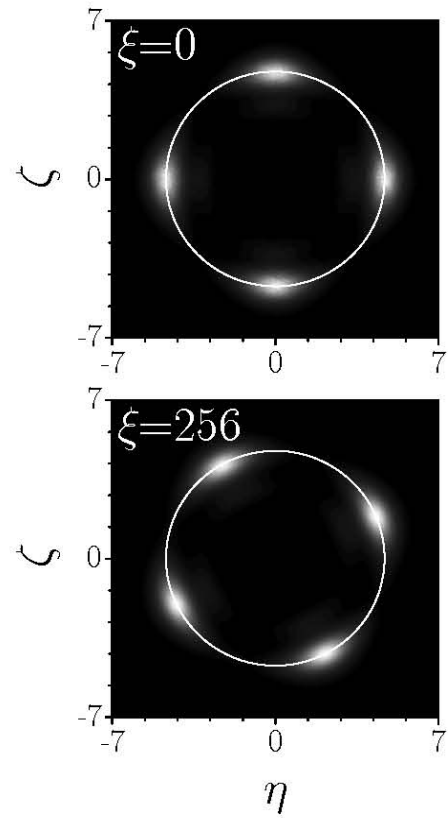

**Figure 6**. Stable propagation of a perturbed quadrupole-mode soliton with frequency $\alpha = 0.1$ at $p = 10$, $\Omega = 2$, and $n = 4$.